%% file: neurips_2024.tex
\documentclass{article}

\usepackage[nonatbib, preprint]{neurips_2024}

\usepackage[utf8]{inputenc} 
\usepackage[T1]{fontenc}    
\usepackage{hyperref}       
\usepackage{url}            
\usepackage{booktabs}       
\usepackage{amsfonts}       
\usepackage{nicefrac}       
\usepackage{microtype}      

\usepackage{float,color}
\usepackage{makecell}
\usepackage{comment} 
\usepackage[normalem]{ulem}
\usepackage{algorithmic} 
\usepackage{algorithm} 
\usepackage{array}
\usepackage{xurl}
\usepackage{caption}
\usepackage{enumitem,kantlipsum}
\usepackage{placeins}
\usepackage{amsmath}
\usepackage[utf8]{inputenc} 
\usepackage[T1]{fontenc}    
\usepackage[pdftex]{graphicx}
\usepackage{grffile}
\usepackage{subfigure}
\usepackage{wrapfig}
\usepackage{lipsum}
\usepackage{subcaption}

\newtheorem{pro-stat}{Problem Definition}

\usepackage[most]{tcolorbox}
\definecolor{block-gray}{gray}{0.85}
\newtcolorbox{myquote}{colframe=black,boxrule=1pt,
colback=white,grow to right by=-1mm,grow to left by=-1mm,
boxsep=0pt,breakable}

\newtcolorbox{inside_myquote}{boxrule=0pt,
colback=block-gray,grow to right by=3mm,grow to left by=3mm,
top=0pt,bottom=0pt}

\usepackage{times}
\usepackage{latexsym}
\usepackage{algorithmic}  
\usepackage{amsmath}  
\usepackage{algorithm} 
\usepackage[algo2e]{algorithm2e} 
\usepackage[utf8]{inputenc} 
\usepackage[T1]{fontenc}    
\usepackage{hyperref}       
\usepackage{url}            
\usepackage{booktabs}       
\usepackage{amsfonts}       
\usepackage{nicefrac}       
\usepackage{microtype}      
\usepackage{xcolor,colortbl}
\usepackage{pifont,dsfont}
\usepackage{wrapfig,lipsum,booktabs}
\usepackage[export]{adjustbox}

\usepackage{soul}
\usepackage{amsthm}

\newcommand{\myModel}{\textsc{LQOT}}

\title{Logic Query of Thoughts: Guiding Large Language Models to Answer Complex Logic Queries with Knowledge Graphs\\ 
}

\author{ Lihui Liu, Zihao Wang, Ruizhong Qiu, Yikun Ban, Eunice Chan, Yangqiu Song \\
\textbf{Jingrui He, Hanghang Tong }\\
  University of Illinois Urbana-Champaign \\
  Computer Science Department \\
  The Hong Kong University of Science and Technology \\
  Computer Science Department \\
  \texttt{lihuil2, rq5, yikunb2, ecchan2, jingrui, htong@illinois.edu} \\
   \texttt{zwanggc@connect.ust.hk, yqsong@cse.ust.hk}
 }

\begin{document}

\maketitle

\begin{abstract}
\input{000abs.tex}

\end{abstract}

\vspace{-0.8\baselineskip}
\section{Introduction}

\input{001intro.tex}

\section{Problem Definition and Preliminaries}\label{problem-definition}

\input{002probl.tex}

\section{Proposed Method}\label{overview}

\input{003framework.tex}\label{sec:method}

\section{Experiments}\label{experiments}

\input{005experiments.tex}

\section{Related work}\label{related-work}

\input{006related_work.tex}

\section{Conclusion}\label{conclusion}

\input{007conclusion.tex}

\bibliographystyle{unsrt}
\bibliography{008reference.bib}

\clearpage

\appendix

\input{009appendix.tex}

\end{document}

%% file: 000abs.tex
Despite the superb performance in many tasks, 
large language models (LLMs) bear the risk of generating hallucination or even wrong answers when confronted with tasks that demand the accuracy of knowledge. 
The issue becomes even more noticeable when addressing logic queries that require multiple logic reasoning steps.
On the other hand, knowledge graph (KG) based question answering methods are capable of accurately identifying the correct answers with the help of knowledge graph, yet its accuracy could quickly deteriorate when the knowledge graph itself is sparse and incomplete. 
It remains a critical challenge on how to integrate knowledge graph reasoning with LLMs in a mutually beneficial way so as to mitigate both the hallucination problem of LLMs as well as the incompleteness issue of knowledge graphs. 
In this paper, we propose `Logic-Query-of-Thoughts' (\myModel) which combines LLMs with knowledge graph based logic query reasoning.
~\myModel\ seamlessly combines knowledge graph reasoning and LLMs, effectively breaking down complex logic queries into easy to answer subquestions. 
Through the utilization of both knowledge graph reasoning and LLMs, it successfully derives answers for each subquestion. By aggregating these results and selecting the highest quality candidate answers for each step, ~\myModel\ achieves accurate results to complex questions. Our experimental findings demonstrate substantial performance enhancements, with 
up to 20\%  improvement over ChatGPT. The code can be found at \url{https://github.com/lihuiliullh/LGOT/tree/main}

%% file: 001intro.tex
Large language models (LLMs) have exhibited remarkable performance across various natural language processing (NLP) tasks, including question answering~\cite{gpt3}, machine translation~\cite{bahdanau2016neural}, text generation~\cite{txt_generation}, recommender system ~\cite{recommendation} and so on. Recent increases in the model size have further enhanced the learning capabilities of LLMs, propelling their performance to new heights. Notable examples of these advanced models include GPT-4~\cite{openai2023gpt4} and Llama 2~\cite{llama2}, among others.

Despite the great success in various domains, LLMs have faced criticism for their limited factual knowledge. Specifically, LLMs tend to memorize facts and knowledge present in their training data~\cite{petroni2019language}. However, research has revealed that LLMs struggle with factual recall and could generate factually incorrect statements, leading to hallucinations~\cite{pan2023unifying}. For instance, when asked, `{\tt When did Einstein discover gravity?}', LLMs might erroneously respond with `{\tt 1687},' which contradicts the fact that Isaac Newton formulated the theory of gravity ~\cite{pan2023unifying}. This issue significantly undermines the reliability of LLMs, especially in contexts requiring multi-step reasoning.
Efforts have been made to address these limitations using approaches like Chain-of-Thought ~\cite{wei2023chainofthought}, but our experiments reveal that even these methods still fall short of mitigating these issues.

Different from Large Language Models (LLMs), knowledge graphs store structured human knowledge, making them a valuable resource for finding answers. Knowledge Graph Question Answering (KGQA) aims to identify an answer entity within the knowledge graph to respond to a given question. Compared with LLMs, KGQA generates more accurate results when the knowledge graph is complete. 
However, the performance of KGQA deteriorates quickly when the underlying KG itself is incomplete with missing relations.
A natural question arises: {\em how can we combine knowledge graph question answering with LLMs so that they can mutually benefit each other?}

In this paper, we introduce a novel model that combines Large Language Models (LLMs) with knowledge graph reasoning to enhance question-answering capabilities. Drawing inspiration from the concepts of Chain-of-Thoughts ~\cite{wei2023chainofthought}, Tree-of-Thoughts ~\cite{yao2023tree} and Graph-of-Thoughts ~\cite{besta2023graph}, we present `Logic-Query-of-Thoughts' (\myModel) which answers complex logic queries by reasoning on multiple subquestions. 
Different from CoT, ToT and GoT which generate thoughts automatically, \myModel\ incorporates a logical query structure and human-defined logic transformations to guide LLMs and knowledge graph reasoning algorithm in finding answers. 
More precisely, \myModel\ utilizes a combination of knowledge graph reasoning methods and LLMs to identify potentially correct answers for each subquestion. These results are merged to generate a comprehensive answer set. Then, the merged answers serve as input for subsequent subquestions. 
This iterative process continues until the final answers are obtained. Our experimental results demonstrate that the proposed \myModel\ significantly enhances the performance, with up to 20\% improvement over ChatGPT.

The main contributions of this paper are: 
(1) {\bf Problem:} we demonstrate that LLMs are insufficient or even deficient at answering complex logic queries which need multiple reasoning steps.
(2) {\bf Algorithm:} we introduce the concept of `Logic-Query-of-Thoughts' (\myModel).
\myModel\ answers complex logic queries by decomposing it to multiple subquestions and guides LLMs in accordance with the logic query.
(3) \textbf{Evaluations:} experimental results on various real-world datasets consistently highlight the state-of-the-art performance achieved by the proposed {\myModel}. 

The code and datasets will be made publicly available upon the acceptance of the paper. 


%% file: 002probl.tex

A knowledge graph can be denoted as $\mathcal{G}=(\mathcal{E}, \mathcal{R}, \mathcal{L})$ where $\mathcal{E} = \{e_1, e_2, ..., e_n\}$ is the set of nodes/entities, $\mathcal{R} = \{r_1, r_2, ..., r_m\}$ is the set of relations and $\mathcal{L}$ is the list of triples.
Each triple in the knowledge graph can be denoted as $(h, r, t)$ where $h \in \mathcal{E}$ is the head (i.e., subject) of the triple, $t \in \mathcal{E}$ is the tail (i.e., object) of the triple and $r \in \mathcal{R}$ is the edge (i.e., relation, predicate) of the triple which connects the head $h$ to the tail $t$. 

Complex logical queries, or Existential First-Order (EFO) queries studied in most literature~\cite{wang2021benchmarking,yin2024rethinking}, can be represented with First-Order Logic (FOL) formula that includes existential quantifier ($\exists$), conjunction ($\wedge$), disjunction ($\vee$) and negation ($\neg$) .
Let $v_1, ..., v_k$ be existentially quantified variables, and $v_?$ be the target variable. An EFO query is written as
\begin{align*}
q[v_?] = v_?. \exists v_1, ..., v_k : \pi_1 \lor \pi_2 \lor ... \lor \pi_n,
\end{align*}
where $\pi_i = \varrho^i_1 \land \varrho^i_2 \land ... \land \varrho^i_m$ is a conjunctive query. 
Each $\varrho^i_m$ denotes a logical atom formula or its negation on relation $r$ between an entity and a variable or two variables. That is, it can be one of $r(u, v)$, $\neg r(u, v)$, $r(v_1, v_2)$ or $\neg r(v_1, v_2)$, where $u\in\mathcal{E}$ is an entity.

When answering complex logical queries, we can directly utilize language models to answer them, or in many cases~\cite{yin2024rethinking}, we can transform the complex logical query as a computation graph~\cite{ren2020query2box} with operator nodes of projection, intersection, negation, and union.
This directed graph demonstrates the computation process to answer the query. Each node of the computation graph represents a distribution over a set of entities in the KG, and each edge represents a logical transformation of this distribution. 
The mapping along each edge applies a certain logical operator:

\noindent\textbf{Projection operation.} Given an input set  $S_{\rm in} \subseteq \mathcal{E}$ and relation $r \in \mathcal{R}$, the projection operation obtains $S_{\rm out} = \cup_{e \in S} Ar(e)$, where $Ar(e) \equiv \{e_i \in \mathcal{E} : r(e, e_i) = \text{True}\}$.

\noindent\textbf{Intersection operation.} Given two input sets $S_{{\rm in}, 1}$ and $S_{{\rm in }, 2}$, the intersection operation outputs $S_{\rm out} = S_{{\rm in}, 1} \cap S_{{\rm in}, 2}$.

\textbf{Negation operation.} Given a set of entities \( S \subseteq \mathcal{E} \), the negation operation computes its complement \( \overline{S} \equiv \mathcal{E} \setminus S \)

\textbf{Union operation.} Given a set of entities \( S_1 \subseteq \mathcal{E} \) and \( S_2 \subseteq \mathcal{E} \), the union operation computes their union \( {S} \equiv S_1  \cup S_2 \)

\subsection{Fuzzy Logic}

Different from Boolean logic, fuzzy logic assigns a value in  $[0, 1]$ to every logical formula. 
A t-norm $\top : [0, 1] \times [0, 1] \to [0, 1]$ represents generalized conjunction in fuzzy logic. Analogously, t-conorms are dual to t-norms for disjunction in fuzzy logic, the t-conorm is defined as $\bot(x, y) = 1 - \top(1 - x, 1 - y)$.   
The negator is $n(x) = 1 - x$. When applying fuzzy logic to FOL queries, following ~\cite{bai2023answering, CQD, chen2022fuzzy}, the complex logical query answering problem can be formulated as 
\begin{align*}\small
q[v?] = v?. \quad v_1, \ldots, v_N = \arg\max_{v_1, \ldots, v_N \in \mathcal{E}} \left( (\varrho^1_1 \top \ldots \top \varrho^1_{m_1}) \bot \ldots \bot (\varrho^n_1 \top \ldots \top \varrho^n_{m_n}) \right) \\
\end{align*}
where $\varrho^i_j \in [0, 1]$ is scored by a pretrained KGE model according to the likelihood of the atomic formula.  In this paper, we use the product t-norm, where given $a, b \in [0, 1]$: 
$
\top(a, b) = a \cdot b, \quad \bot(a, b) = 1 - (1 - a)(1 - b).
$

\subsection{Problem Definition}

When answering complex logical queries, finding the answers directly in the knowledge graph is hindered by its incompleteness. On the other hand, using LLMs to answer complex logical queries may result in incorrect answers due to the hallucination problem. This paper aims to study the integration of LLMs and fuzzy logic reasoning to effectively answer logical queries by leveraging the advantages of both approaches. Formally, the problem is defined as follows.

\textbf{Given:} (1) A knowledge graph $\mathcal{G}$, (2) a complex logical query, (3) a large language model, (4) a fuzzy logic reasoning model;
\textbf{Output:} (1) the answer of the question.


\begin{figure}
	\centering
    \includegraphics[width=0.45\textwidth]{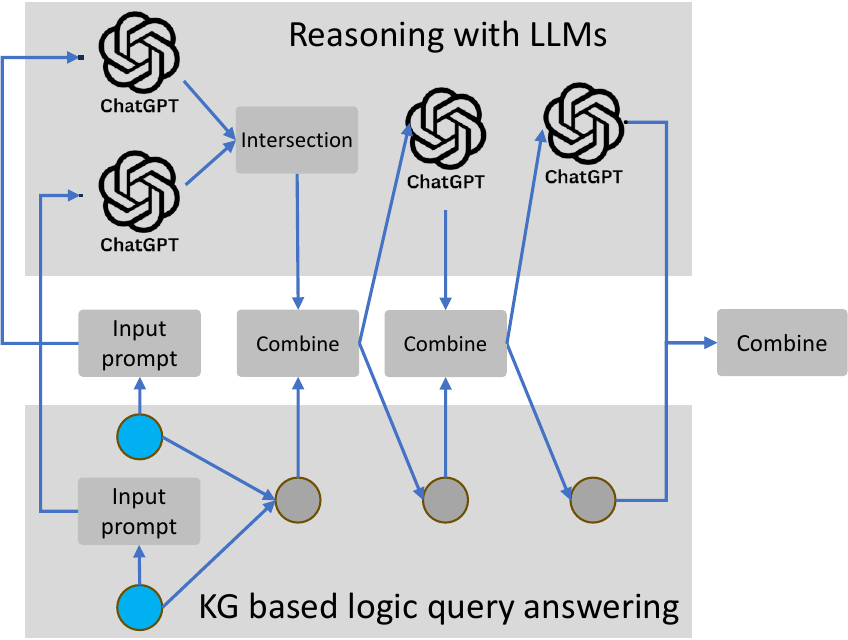}
	\caption{Framework of the Proposed LQOT.}
	\label{lgot}
 \vspace{-1.8\baselineskip}
\end{figure}

%% file: 003framework.tex
In this paper, we design a model, \myModel, to answer complex logical queries by combining LLM and KGQA (We use KGQA to denote finding answers in the knowledge graph through fuzzy logic reasoning), in a mutually beneficial way to leverage the advantages of both. Because large language models are not good at answering complex logical queries, \myModel\ decomposes the original complex logical query into multiple easy-to-answer subquestions according to the atom predicates in its dependency graph and executes each subquestion in their corresponding order in the computation graph. For each atom predicate, both the large language model and fuzzy logic reasoning method are used to solve it according to its corresponding logical operator. Figure~\ref{lgot} introduces the framework of \myModel, where the same operator is executed by both the LLMs and KGQA, and \myModel\ combines the input and output.

Our proposed method has several advantages compared with existing methods. Previous retrieval-augmented generation-based methods, e.g., ~\cite{choudhary2024complex}, usually fail to find the correct answers due to the incompleteness of the knowledge graph. Therefore, we let the language model call the fuzzy logic reasoning method and use its answers to augment the LLM's answers. At the same time, the answers generated by the LLMs for each atomic question are also used to augment the results of the fuzzy logic reasoning method in case the knowledge graph is incomplete. 
This ensures that these two models can mutually assist each other. 

We chose fuzzy logic reasoning due to several reasons, first, fuzzy vector can handle different types of logical operations, and all these operations are closed. Second, fuzzy vector can effectively connect LLMs with fuzzy logic reasoning model since we only need to update the value of the fuzzy vector according to the results of LLMs, which is much more convenient compared to relearning the box embedding ~\cite{ren2020query2box,newlook} or probability distribution ~\cite{ren2020beta}.
The details of the method are introduced in the following sections.

\subsection{Question Transformation for LLMs}

Answering complex logical queries requires multiple reasoning steps. To address this, we divide the original hard-to-answer question into multiple easy-to-answer subquestions, which are executed in order and ultimately lead to the final results. More specifically, for each atom predicate in the dependency graph, we first translate it into a natural language question and then use large language models to find its answer. We employ the following approach to transform the atomic relation $r$ in the logical query into its corresponding question prompt.

Given an atom predicate $r$, let $S_r = \{(h_i, t_i) | (h_i, r, t_i) \in G\}$, which comprises of all entity pairs associated with $r$ in the knowledge graph. Our objective is to formulate a query that facilitates the retrieval of $t_i$ given both $h_i$ and $r$. For this purpose, we utilize the following prompt to transform $r$ to a question prompt:

\begin{myquote}

Prompt: Atom Predicate to Question

\begin{inside_myquote}\small
Given entity pairs \{text\_A\}. 
The relationship between the first entity and the second entity in the pair is \{text\_B\}. Please rewrite the relationship to a question of entity1, so that the answer is entity2. 
\end{inside_myquote}

\end{myquote}

\subsection{Logical Operators}

So far, we have described how to transform each atom predicate in the dependency graph into a natural language question. Now, we will describe how to use both large language models and fuzzy logic reasoning models to find answers. To answer a query according to the computation graph, we need to define logical operators for both the language models and the fuzzy logic reasoning method. In the following section, we describe the design of the different logical operators used in the computation graph.

\noindent\textbf{Projection operation:} given an entity set $S_1$, the goal of projection operation is to find answer set $S_2$ which is connected to any entities in $S_1$ with relation $r$ in any case. 

For large language models, 
the projection operation takes a set of entities and a question prompt as input and 
generates an additional set of answers. 
To achieve this with LLMs, we employ the following prompt construction method: given a set of input entities $S$, we concatenate them using the word "{\tt or}". For example, if $S = \{{\tt Tom Hanks}, {\tt Tom Cruise}\}$, the resulting prompt would be `{\tt Tom Hanks or Tom Cruise}.'
Each atom predicate corresponds to a question prompt defined above, and we construct the input prompt by combining the question prompt with the relevant entities. 
Here is an example: Suppose we have the relation $(h, {\tt isDirectedBy}, ?)$ with its question prompt `{\tt which movies are directed by h?}' In this case, the input prompt would be `{\tt which movies are directed by Tom Hanks or Tom Cruise?}'. 
The detailed prompt for the Projection Transformation is provided in Appendix.

For fuzzy logic reasoning, we represent each variable in the computation graph as a fuzzy vector $T \in \mathbb{R}^{|\mathcal{E}|}$~\cite{CQD}, where the $i$-th element in the fuzzy vector indicates the probability that entity $e_i$ is the correct answer for the current variable node.
We define an operator $M_r \in [0, 1]^{|\mathcal{E}| \times |\mathcal{E}|}$ for each relation $r$, where $M_r(i,j)$ denotes the probability that $e_i$ is connected with $e_j$ by relation $r$. We utilize knowledge graph embedding methods $g_{r_i}(e_i, e_j)$ to score the likelihood that $e_i$ is connected with $e_j$ by relation $r_i$, where $g()$ is calculated based on the embeddings of $e_i$ and $e_j$. Many KGE methods can be used as $g()$, in this paper, we adopt ComplEx ~\cite{complEx} which is widely used in many applications.
The details of building $M_r$ can be found in the Appendix. Note that because the knowledge graph is very sparse, $M_r$ can be stored in a single GPU, and the reasoning process is efficient. 
Let $\max_i$ and $\max_j$ denote the maximum over rows and columns of a matrix, respectively. Then the projection operation for ($v_k, r, v_?$) would be
$
T(v?) = \max_j ((T(v_k)^\top \cdots \times |\mathcal{E}|) \odot M_r)
$.
If $v_k$ is a constant entity $c = e_i$, then
$
T(v?) = \text{row}_i(M_r)
$

\textbf{Intersection operation:}
give the answer sets of different subquestions, the intersection operator is used to get the answers that belongs to all question. 
For large language models, the intersection operation can be simplly modeled as the intersection of two answer sets.
When reasoning with fuzzy vector,
if a node $v?$ is formed by merging child nodes $\{v_1, \ldots, v_K\}$ through intersection, then the intersection can be calculated by: 
$T(v?) = \prod_{1 \leq i \leq K} T(v_{i})$.

\noindent\textbf{Negation operation}: given a set of entities, the negation contains all entities that are not belong to this set. 
The negation for large language model is all entities in the knowledge graph which do not belong to the answer of LLMs. 
For fuzzy logic reasoning, 
the negation operation for $\neg r(v_k, v_?$) would be
$
T(v?) = \max_j ((T(v_k)^\top \cdots \times |\mathcal{E}|) \odot (\mathbf I - M_r))
$.
If $v_k$ is a constant entity $c = e_i$, then
$
T(v?) = \text{row}_i(\mathbf I -M_r)
$
.

\noindent\textbf{Union operation}: 
The union operation of LLMs is the union of their answers. For fuzzy logic reasoning, if a node $v?$ is formed by merging child nodes $\{v_1, \ldots, v_K\}$ through union, then the union can be calculated by: 
$T(v?) = \mathbf I - \prod_{1 \leq i \leq K} (\mathbf I - T(v_{i}))$.

\subsection{Combination of LLMs and Fuzzy Logic Reasoning}

When modeling the operations with LLMs, we can translate the input information into input prompts and get the output by parsing the LLM outputs. Given that the LLMs are generative models, the output set is assumed to be a very limited set of entities with estimated confidence values: 
$$S_{\rm out}^{\rm LLM} = \{(e, p(e))\},$$
where $|S_{\rm out}^{\rm LLM}|=K\ll|\mathcal{E}|$, $p(e)$ is the confidence for the entity $e$. We note that estimating $p(e)$ is non-trivial, and $p(e)$ can be zero in most cases (e.g., LLMs fail to output all the correct answers), so how to estimate $p(e)$ is an important task. 

On the other hand, 
current logical query answering methods primarily employ retrieval-based approaches. In these methods, scores are assigned to all entities in $\mathcal{E}$ to indicate the likelihood that each entity is the correct answer. 
$$S_{\rm out}^{\rm KGQA} = \{(e, {\rm Score}(e)): e\in \mathcal{E}\},$$
where $|S_{\rm out}^{\rm KGQA}| = |\mathcal{E}|$. 
KGQA methods rank all existing entities based on their scores, which we assume to be within the range of 0 to 1 for convenience.

The success of \myModel\ heavily relies on combining the advantages of LLMs and fuzzy logic reasoning to mitigate their disadvantages. For LLMs, the outputs are featured with high accuracy of a small set of predictions but might suffer from randomness and even hallucination. 
For fuzzy logic reasoning, the outputs may suffer from knowledge graph incompleteness. We discuss how to obtain the final answer by obtaining the outputs $S_{\rm out}^{\rm KGQA}$ and $S_{\rm out}^{\rm LLM}$.

\noindent\textbf{Likelihood ratio test for LLM outputs.}
Inspired by the likelihood ratio test ~\cite{king1998unifying}, we use a likelihood ratio-based scheme to 
select potential correct answers from LLMs. Let $p(e_1),\dots,p(e_N)$ denote the predicted likelihoods of the answers, and let $p_{\max}:=\max_{i=1}^N p(e_i)$.
Ideally, if there are $K$ correct answers, a perfect LLM should predict $p(e_i)\approx\frac1K\approx p_{\max}$ for the correct answers, and $p(e_i)\approx 0$ for the wrong answers. Since $K$ is unknown to us, we cannot directly decide their correctness via $p(e_i)$.
Instead, we decide the correctness of each answer $e_i$ through the \emph{likelihood ratio} $\frac{p(e_i)}{p_{\max}}$. For each answer $e_i$, the likelihood ratio should be
\begin{equation}
\frac{p(e_i)}{p_{\max}}\approx\begin{cases}
1,&\text{if }e_i\text{ is correct};\\
0,&\text{if }e_i\text{ is wrong}.
\end{cases}
\end{equation}
Since the likelihood ratio $\frac{p(e_i)}{p_{\max}}$ does not depend on $K$, it allows us to decide their correctness without knowing the number of correct answers. 

\noindent\textbf{Enhancing KGQA with LLM.}
Since knowledge graphs are often incomplete, KGQA methods may encounter issues due to this incompleteness. To address this problem, we select answers with scores $p(e_i)/p_{\max}\ge\theta$ identified by LLMs. We then assign the score $\alpha * p(e_i)$ to the fuzzy vector obtained by the fuzzy logic reasoning method, thereby completing the fuzzy vector.

\noindent\textbf{Enhancing LLM with KGQA.} 
Answers generated by LLMs may be incorrect due to the hallucination problem. However, answers from fuzzy logic reasoning are more likely to be correct if they have a high score ${\rm Score}()$. Therefore, after updating the fuzzy vector, we select the top-$\min\{k,10\}$ answers based on their scores and consider them as the correct answers to refine LLMs' output.

\subsubsection{Answer Evaluation}
After obtaining answers by combining LLMs with KGQA, we can further utilize an answer evaluator to refine the selection of answers. The purpose of the answer evaluator is to leverage Large Language Models (LLMs) to assess the quality of generated responses. Previous research has shown that LLMs are proficient at selecting the most appropriate answers from multiple choices, rather than directly generating answers~\cite{robinson2023leveraging}.

For a given vertex $v$ in the logic query, whether it represents an intermediate or final answer, answer evaluation is employed to gauge the likelihood that an entity $e_i$ is the correct answer for $v$. More formally, when presented with a set of entities $S \subseteq \mathcal{E}$, the answer evaluation process selects a subset of entities from $S$ that are likely to be correct answers for $v$.
The prompt for the evaluation process is shown below.
It is important to note that this module is optional.

\begin{myquote}
Prompt: Answer Evaluation

\begin{inside_myquote}\small
Replace the answers with correct answers if the answers in the choices are not correct.
Given the question \{text\_A\} and its potential answer choices \{text\_B\}, output top10 answers in a json format. The output json has key "input", "answer". The answers don't need to be in the potential answer choices.
\end{inside_myquote}

\end{myquote}

An example of prompts used in \myModel\ can be found in Figure ~\ref{prompt} in Appendix.

\subsection{Time Complexity}

The computational cost of the proposed method primarily arises from querying large language models (LLMs). The knowledge graph (KG) reasoning process is highly efficient, with its time consumption being negligible. Similarly, the combination step—integrating the outputs of the LLMs with those from KG reasoning—is also very fast and can be ignored in terms of time complexity.
In essence, the time complexity of the proposed method aligns with that of most LLM-based reasoning approaches. The KG reasoning and combination steps can be considered as external tool calls made by the LLMs.

%% file: 005experiments.tex

\subsection{Experimental Setting}

Existing logical query reasoning methods commonly test on dataset FB15k, FB15k-237, NELL995 and so on. One problem for these datasets is that their nodes and edges are represented as meaningless id which contains no semantic meaning. This seriously hinder the ability of large language models. 
To solve this problem, we use three different datasets: 
(1) \textbf{MetaQA} ~\cite{embedkgqa} is a multi-hop question dataset in the movie domain. 
The original MetaQA dataset only contains three types of questions, which are 1-hop, 2-hop and 3-hop questions. To support negation and union, we create new queries 2u and pin based on existing queries. In the experiment, we randomly sample a subset of the test queries to save money.  
(2) \textbf{ComplexWebQuestions} ~\cite{lan:acl2020} contains the 1-hop, 2-hop and 3-hop questions. We only use the 3-hop questions in the experiment. 
(3) \textbf{GraphQuestions} ~\cite{graphQuestions} is a QA dataset consisting of a set of factoid questions with logical forms and ground-truth answers. We use 2-hop questions and 2 intersection questions in the experiment. 
All questions in ComplexWebQuestions and GraphQuestions datasets can be answered with Freebase.
Because the original query sets is very large, so we only use a subset of them. 
The statistics of the background knowledge graphs are shown in Table ~\ref{datasets}. 
In the experiment, we compare the proposed \myModel\ with baselines on an incomplete KG with 50\% missing edges (we randomly delete 50\% of the edges from the full knowledge graph). When the knowledge graph is complete, traversing the graph alone can achieve 100\% accuracy. \textbf{Note that all these datasets are complex logical queries.}

Six baselines are used in the experiments, including
(1) ChatGPT ~\cite{gpt3}. 
(2) Chain-of-Thought with ChatGPT ~\cite{wei2023chainofthought}.
(3) Query2Box \cite{ren2020query2box}: a recent knowledge representation model that encodes logical queries as box embeddings.
(4) BetaE ~\cite{ren2020beta}: a knowledge representation model, which utilizes the beta distribution to encapsulate logical queries.
(5) QTO ~\cite{bai2023answering}: the state-of-the-art complex logical query answering method. 
(6) LARK ~\cite{choudhary2024complex}: a method that uses LLMs to answer complex logical queries. We use Llama-2-7b in the experiment. 

We adopt Hit ratio at $k$ (Hit@$k$) as the metric to measure the performance of different baselines. 
Hit ratio at $k$ (Hit@k) is the fraction of times a correct answer was retrieved within the top-$k$ positions. We use $k=1,3$, and 10 in the experiments. 

\begin{figure*}
	\centering
	\includegraphics[width=0.85\textwidth]{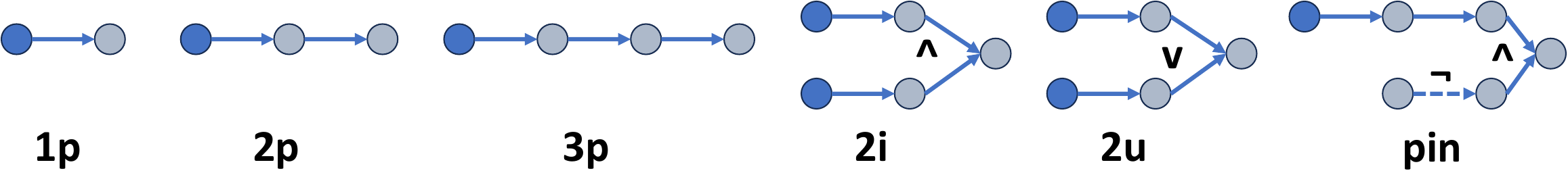}
	\caption{Query structures, illustrated in their query computation tree representations.}
	\label{query_type}
 \vspace{-1\baselineskip}
\end{figure*}

\begin{table*}
	\centering
	\caption{Question answering results of MetaQA on 50\% incomplete knowledge graphs.}
 \setlength\tabcolsep{0.8pt}
 \scalebox{0.75}{
	\begin{tabular}{|c|c|c|c|c|c|c|c|c|c|c|c|c|c|c|c|}
	\hline
	Dataset         & \multicolumn{3}{c|}{1p} & \multicolumn{3}{c|}{2p} & \multicolumn{3}{c|}{3p}  & 
    \multicolumn{3}{c|}{2u} & 
    \multicolumn{3}{c|}{pin} \\ \hline
        Metric         & Hit@1 & Hit@3  & Hit@10 & Hit@1 & Hit@3  & Hit@10 & Hit@1 & Hit@3  & Hit@10 & Hit@1 & Hit@3  & Hit@10 & Hit@1 & Hit@3  & Hit@10 \\ \hline
       \multicolumn{16}{|c|}{Only KG Reasoning} \\ \hline
	Query2Box &  19.7  &  57.5  & 72.7 & 14.1  & 44.7 & 70.1 & 20.1 & 33.7 &  47.7 &  70.4 & 75.3 & 77.8  & 36.1 &  40.5 & 45.5
   \\ \hline
	BetaE   &  10.2  &  16.6  & 24.8 & 10.9  & 16.4 & 23.6 &  27.8 &  34.1 & 44.9 & 72.8 & 74.4  & 82.3 & 33.1 & 39.5 & 44.5
 \\ \hline
    QTO & 61.4  &  62.8    &  65.7 & 40.9  & 44.6 & 49.4 &  44.9 & 50.5 &   57.9 & 86.4 & 91.8 & 93.3 & 48.1 & 56.5 & 57.5 \\ \hline
    \multicolumn{16}{|c|}{LLMs} \\ \hline
	ChatGPT & 66.1 &    80.3   &  84.3 &  45.4  & 57.0 &  64.1 &  41.6  &  52.6 &   55.6 & 44.0 & 77.9  & 88.8 & 11.2 & 21.1 & 29.8 \\ \hline
    CoT (ChatGPT) & 64.1 &  69.0 & 69.4 & 35.7 & 44.5 & 45.2 & 40.1 & 45.9 & 46.2 & 42.4 & 71.3  & 82.8 & 10.7 & 18.5 & 22.7 \\ \hline
    GPT4 & 61.9 &  69.5 & 81.3 & 19.2 & 37.5 & 51.5 & 34.9 & 52.6 & 63.3 & 55.7 & 84.1  & 91.0 & 15.4 & 30.9 & 40.1 \\ \hline
    \multicolumn{16}{|c|}{LLMs + KG Reasoning} \\ \hline
        LARK & 6.7 & 56.1 & 63.2 & 3.3 & 28.1 & 53.2  & 2.7 & 26.2 & 56.1 & 0.0 & 85.9 & 89.7 & 0.0 & 0.0 & 0.0 \\ \hline
\rowcolor{lightgray}	\myModel\ &  77.7  & 86.3   & 89.7  &  58.7 & 79.0 & 83.7  &  67.9 & 83.9  &  90.8  & 90.4 & 98.3 & 99.4 & 52.6 & 62.8 & 68.7 \\ \hline

	\end{tabular}
 }
	\label{metaqa}
\end{table*}

\begin{table*}
	\centering
	\caption{Question answering results of GraphQuestions and CWQ on full and 50\% incomplete knowledge graphs.}
 \setlength\tabcolsep{0.8pt}
	\vspace{-0.5\baselineskip}
 \scalebox{0.8}{
	\begin{tabular}{|c|c|c|c|c|c|c|c|c|c|c|c|c|c|c|c|c|c|c|c|c|c|}
	\hline
	Dataset         & \multicolumn{3}{c|}{GraphQuestions full KG} & \multicolumn{3}{c|}{GraphQuestions 50\% KG}  & \multicolumn{3}{c|}{CWQ 3hop full KG} & \multicolumn{3}{c|}{CWQ 3hop 50\% KG}   \\ \hline
        Metric         & Hit@1 & Hit@3  & Hit@10 & Hit@1 & Hit@3  & Hit@10 & Hit@1 & Hit@3  & Hit@10 & Hit@1 & Hit@3  & Hit@10 \\ \hline
    QTO      &  100.0  & 100.0 & 100.0  & 26.8  & 26.8 & 26.8 & 96.5  & 100.0  &  100.0 & 30.1 &  35.1 & 38.8    \\ \hline
    CoT (ChatGPT)    & 18.7 &  19.4  & 19.4 & 18.7 & 19.4 & 19.4 & 13.8 & 20.6 & 37.9   & 13.8 & 20.6 & 37.9 \\ \hline
	ChatGPT      & 18.4  &  19.0 &  21.5 & 28.8  & 30.3 &  33.3  & 13.7 & 20.7 & 34.5   & 13.7 & 20.7 & 34.5   \\ \hline
\rowcolor{lightgray}	\myModel\     &   100.0  & 100.0 & 100.0  &  52.2  &  52.8 & 52.8 & 96.5  & 96.5  &  100.0 &  33.3 &  40.7  &  59.2  \\ \hline
	\end{tabular}
 }
	\label{GraphQuestions}
	\vspace{-1\baselineskip}
\end{table*}


\subsection{Performance of Question Answering}

Table~\ref{metaqa} displays the question-answering performance of all models on the MetaQA dataset. Notably, \myModel\ surpasses all baseline methods when operating with the 50\% incomplete knowledge graph. It demonstrates significant results with a 16\% improvement in Hit@1, 12.8\% in Hit@3, and 14.2\% in Hit@10 compared to ChatGPT. Additionally, when compared to the logic query reasoning method QTO, our proposed approach achieves an even more performance boost, with an average improvement of 17.5\% in Hit@1, 24.4\% in Hit@3, and 26.4\% in Hit@10. We omit the results over complete knowledge graph, because both QTO and \myModel\ achieve near 100\% accuracy. 

One interesting observation is that ChatGPT does not perform well on 1-hop questions. One would expect ChatGPT to excel in answering short questions. However, we have found that ChatGPT is not proficient in responding to uncommon questions. For instance, when asked, ``{\tt What does [Walter Steiner] act in?}" the answers provided by ChatGPT include `{\tt Downhill Racer (1969)}', `{\tt The Great White Hope (1970)}', `{\tt The Other Side of the Mountain (1975)}', `{\tt The Other Side of the Mountain Part 2 (1978)}' and so on.
However, none of these answers is correct. This illustrates ChatGPT's limitations in handling less common questions.

We also test the baselines performance on the other two datasets which are GraphQuestions and ComplexWebQuestions. Table ~\ref{GraphQuestions} shows the results. 
As we can see, the proposed \myModel\ consistently outperforms all baseline methods with a large margin. For example, it can achieve  23.4\% gain in Hit@1, 22.5\% gain in Hit@3 and 19.5\% gain in Hit@10 on GraphQuestions compared with ChatGPT in the 50\% incomplete knowledge graph. 
For ComplexWebQuestions, \myModel\ can achieve 19.6\% gain in Hit@1, 20.0\% gain in Hit@3 and 24.7\% gain in Hit@10 compared with ChatGPT in the 50\% incomplete knowledge graph. 


\subsection{Ablation Study}\label{ablation_study}
\begin{figure}
\label{ablation}
\centering
\begin{minipage}{.28\textwidth}
        \centering
        \includegraphics[width=0.9\linewidth]{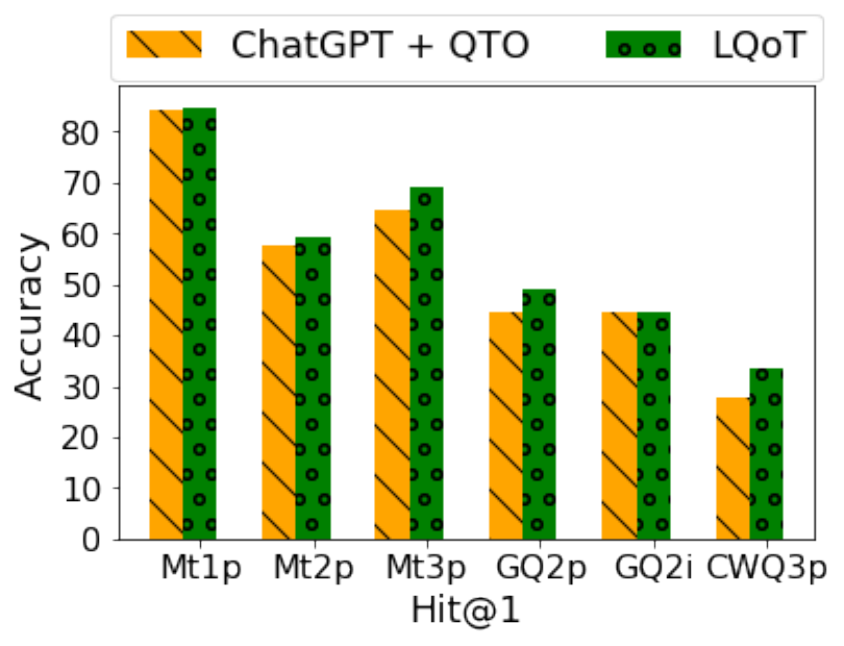}
        \caption{Hit@1 performance.}
        \label{fig:1a}
    \end{minipage}%
\begin{minipage}{0.28\textwidth}
        \centering
        \includegraphics[width=0.9\linewidth]{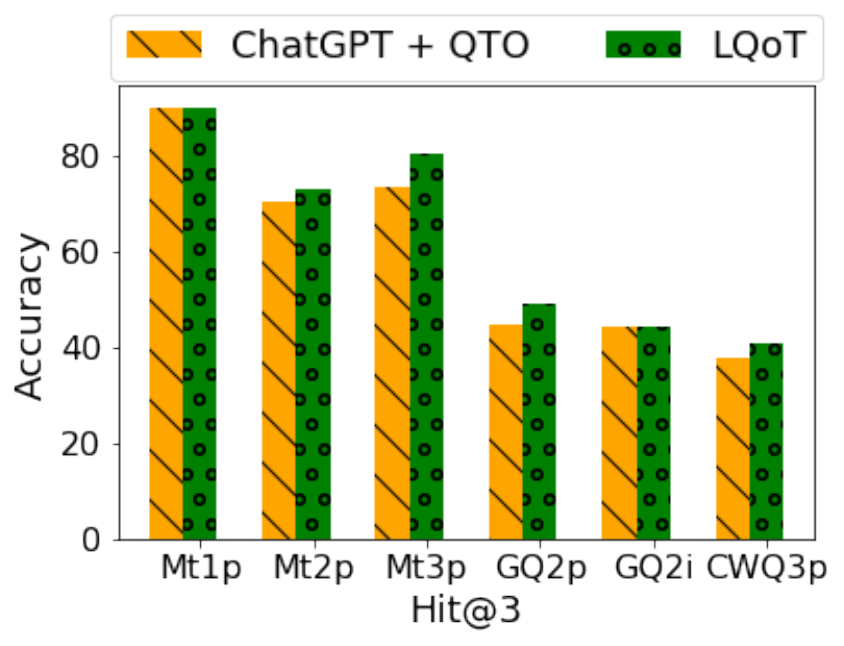}
        \caption{Hit@3 performance.}
        \label{fig:1b}
    \end{minipage}
\begin{minipage}{0.28\textwidth}
        \centering
        \includegraphics[width=0.9\linewidth]{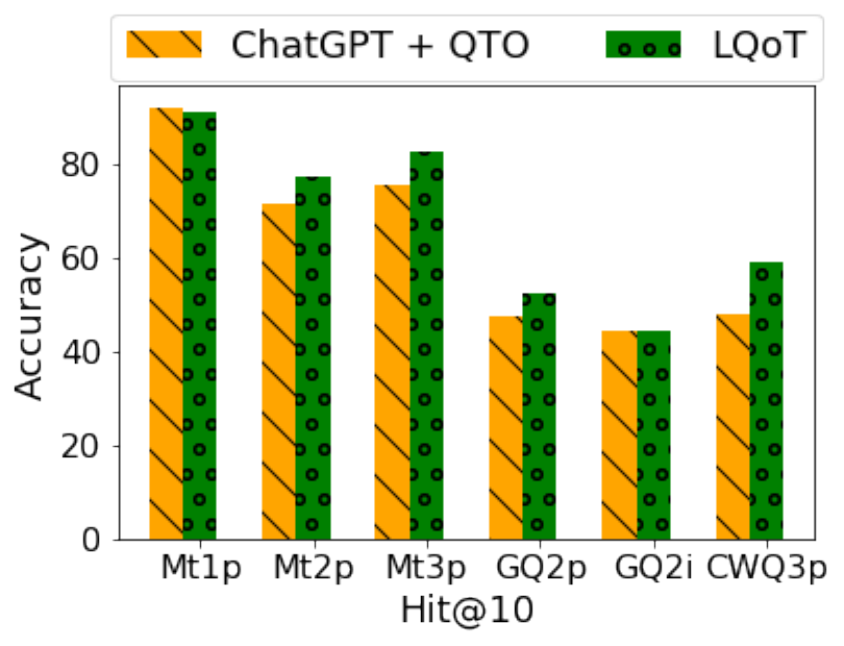}
        \caption{Hit@10 performance.}
        \label{fig:1c}
    \end{minipage}
\label{xxx}
\end{figure}

\noindent{\bf A - ChatGPT + QTO.}
Figure ~\ref{fig:1a} to Figure ~\ref{fig:1c} illustrates the performance when simply combining LLMs with QTO (details can be found in Table~\ref{simple} in Appendix). To elaborate, when given a logic query, we initially use QTO to obtain results, followed by employing ChatGPT to find additional results. These two sets of results are then combined. The results obtained by QTO, assigned with 100\% probability, are placed at the beginning of the answer list, followed by the results from ChatGPT at the end.
The results indicate that \myModel\ achieves an average improvement of 3.26\% in MetaQA 2-hop questions and 6.1\% in MetaQA 3-hop questions. For GraphQuestions, there is a 4.5\% improvement in 2-hop questions and a 6.5\% improvement in ComplexWebQuestions 3-hop questions. These results suggest that as the complexity of logic queries increases (i.e., requiring more reasoning steps), our proposed \myModel\ consistently outperforms the baseline methods.

\noindent{\bf B - The Effectiveness of Answer Evaluation.}
We compare two versions of \myModel: one with Answer Evaluation and one without. The results are displayed in Table~\ref{evaluation}. We employ Answer Evaluation in the ComplexWebQuestions 3-hop dataset, while for other datasets, we do not employ Answer Evaluation. As demonstrated in the table, the inclusion of Answer Evaluation leads to an average performance improvements of 6.2\% for ComplexWebQuestions 3-hop questions. These results show the effectiveness of the proposed Answer Evaluation method.

\begin{table}
	\centering
	\caption{Ablation Study of Answer Evaluation.}
	\setlength\tabcolsep{2pt}
 \scalebox{0.8}{
	\begin{tabular}{|c|c|c|c|c|c|c|}
	\hline
	Model          & \multicolumn{3}{c|}{\myModel\ no Evaluation}  & \multicolumn{3}{c|}{\myModel}  \\ \hline
    \multicolumn{1}{|c|}{} & \multicolumn{6}{c|}{50\% KG} \\ \hline
  Metric         & Hit@1 & Hit@3  & Hit@10 & Hit@1 & Hit@3  & Hit@10  \\ \hline
    CWQ 3hop  & 29.6 &  33.3  & 51.8 & \cellcolor{lightgray}33.3 &  \cellcolor{lightgray}40.7  &  \cellcolor{lightgray}59.2  \\ \hline
	\end{tabular}
 }
	\label{evaluation}
	\vspace{-2\baselineskip}
\end{table}

%% file: 006related_work.tex


\subsection{Large Language Models}
The landscape of large language models has evolved significantly with the development of diverse and powerful models. 
Among these models, GPT-2 ~\cite{gpt2} and GPT-3 ~\cite{gpt3} have gained immense popularity for their remarkable text generation capabilities and the ability to perform a wide range of tasks. 
LLaMA ~\cite{touvron2023llama} and LLaMA2 ~\cite{llama2} are state-of-the-art LLMs capable of generating code, and natural language about code, from both code and natural language prompts.
BARD ~\cite{manyika2023overview} is trained with the latest information from the Internet, so it has more data to gather information in real time. 
These diverse large language models represent an exciting spectrum of approaches, offering researchers and practitioners a rich toolkit to address different NLP tasks.

Recent work on using language models for reasoning tasks are mostly based on prompt engineering. For example, in ~\cite{kojima2023large}, Kojima et al. find that adding some simple information `{\tt Let’s think step by step.}' can greatly improve LLMs' performance on a variety of downstream tasks. 
In ~\cite{wei2023chainofthought}, Wei et al. aim to answer questions by breaking them down into intermediate steps. They  prompt the language model with similar examples where an answer is formed step-wise before providing the requested answer.
Based on this idea, Yao et al. ~\cite{yao2023tree} sample multiple thoughts at each intermediate step and make the model perform deliberate decision making by considering multiple different reasoning paths.
Besta et al. ~\cite{besta2023graph} further generalize this idea and model the information generated by an LLM as an arbitrary graph, where units of information (`LLM thoughts') are vertices, and edges correspond to dependencies between these vertices.
Some other methods, such as ~\cite{zhou2023leasttomost, creswell2022selectioninference}, propose different methods for prompt engineering. 
Some research efforts aim to combine knowledge graphs with large language models (LLMs) to answer natural language questions. Examples include ROG ~\cite{luo2024reasoning} and TOG ~\cite{sun2024thinkongraph}. However, these approaches are limited to addressing simple or multi-hop questions. Other approaches utilize retrieval-augmented generation ~\cite{lewis2021retrievalaugmented} methods to tackle complex natural language questions, such as GraphRAG ~\cite{edge2024local} and REPLUG ~\cite{shi2023replug}. Nevertheless, these methods are not specifically designed to handle complex logical queries. The only work focused on answering complex logical queries is KARL ~\cite{choudhary2024complex}, which is used as the baseline in this paper.


\subsection{Logic Query Answering}
Logic query answering aims to answer First-Order (EPFO) queries based on the information in the knowledge graph. Some commonly used logic operations include existential qualifier ($\exists$), conjunction ($\wedge$), and disjunction ($\vee$), while the more general First-Order Logic (FOL) queries also include negation ($\neg$).
Many methods have been proposed recently. For example, in ~\cite{ren2020query2box}, each logic query is embedded as a box in the embedding space, while each entity in the knowledge graph is embedded as a point in the embedding space. In ~\cite{ren2020beta}, both the logic query and the KG entities are represented as a Beta distribution in the embedding space, and KL divergence is used to calculate their distance. In ~\cite{chen2022fuzzy}, fuzzy vector is used to encode a set of entities, while in ~\cite{bai2023answering}, each operation is defined as a matrix multiplication. 
Other methods, like ~\cite{CQD, zhu2022neuralsymbolic, Choudhary_2021, newlook}, use different ideas, such as hyperbolic embedding, graph neural network and knowledge graph embedding to boost the QA performance. 
Different from these methods, we combine the reasoning capabilities of LLMs and knowledge graph based methods, and use the logic query structure to guide the reasoning process of LLMs.



%% file: 007conclusion.tex
In this paper, we introduce a novel model, \myModel, that utilizes Large Language Models (LLMs) in conjunction with logic query reasoning to tackle logic queries effectively. Our model leverages the structure of the logic query to provide guidance to LLMs in generating answers for each step. It utilizes the query graph and the intermediate answers derived from the knowledge graph question answering (KGQA) method during reasoning, and iteratively finds answers until the final results are obtained. Extensive experimental results show that the proposed LGOT approach can significantly enhance performance, 
with up to 20\%  improvement over ChatGPT. 

Studying this problem has broder impacts, it will greatly mitigate the hallucination problem of LLMs. 
The potential limitation of this work is that it is not applicable to answering ID-based complex logical queries, such as those used in Query2Box ~\cite{ren2020query2box}.

\section{Ethical Considerations}

We have thoroughly evaluated potential risks associated with our work and do not anticipate any significant issues. Our framework is intentionally designed to prioritize usability and ease of implementation, thereby reducing barriers for adoption and minimizing operational complexities. Furthermore, it's essential to highlight that our research builds upon an open-source dataset. This ensures transparency, fosters collaboration, and helps address ethical considerations by providing accessibility to the underlying data.

\section{Limitation}

Our dataset exhibits several limitations that warrant consideration. Firstly, our training data is constrained by its limited scope, primarily focusing on specific domains rather than providing comprehensive coverage across diverse topics. This restriction may affect the model's generalizability and performance in addressing queries outside of these predefined domains. Moreover, while our knowledge graph serves as a valuable resource for contextual information, it is essential to acknowledge its incompleteness. Despite its vast size, certain areas within the knowledge graph may lack sufficient data or connections, potentially leading to gaps in the model's understanding and inference capabilities.

%% file: 009appendix.tex
\begin{figure*}
	\centering
	\includegraphics[width=0.8\textwidth]{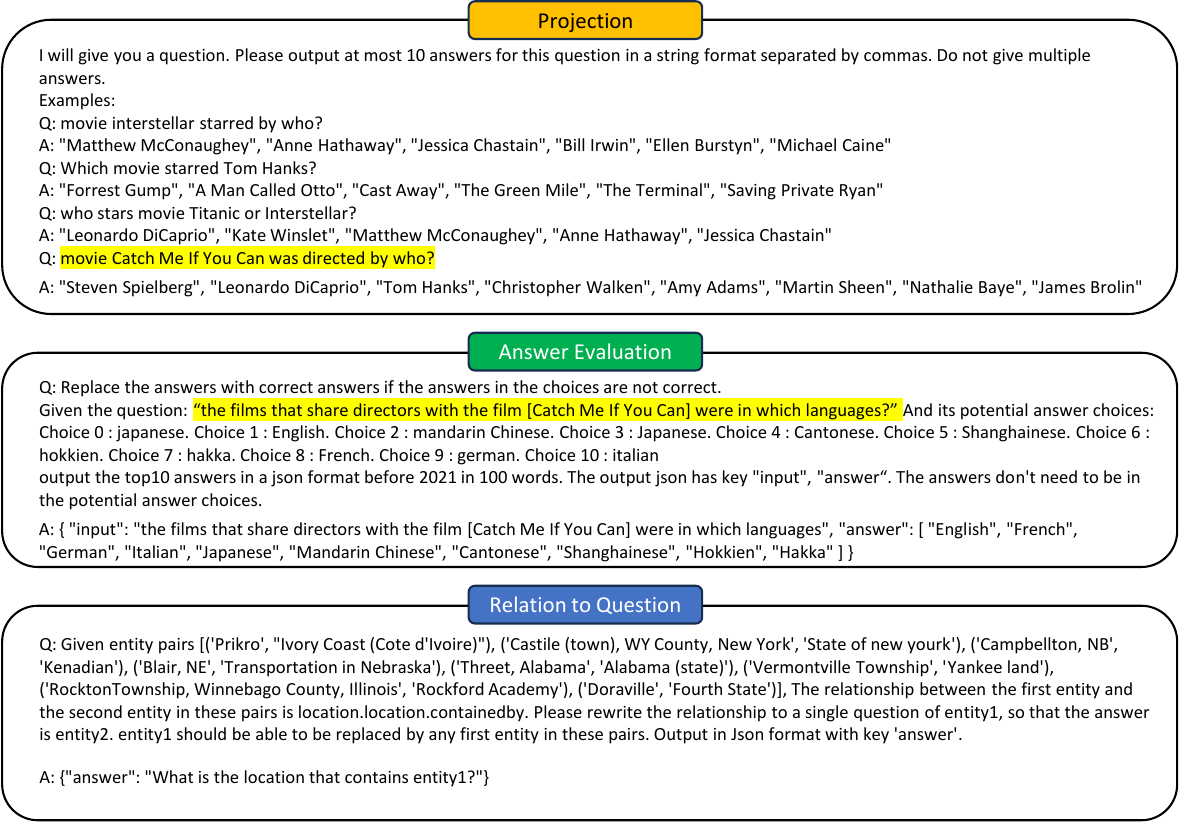}
	\caption{Prompt Examples.}
	\label{prompt}
\end{figure*}

\section{Reproducibility.} All experiments are performed on a machine with an Intel(R) Xeon(R) Gold 6240R CPU, 1510GB memory and NVIDIA-SMI Tesla V100-SXM2. 

\section{Dataset Information}

The details of the background knowledge graphs can be found in Table ~\ref{datasets}. 

\begin{table}[H]
	\centering
	\caption{Statistics of the full and 50\% KG for the three datasets.}
 \scalebox{0.8}{
	\begin{tabular}{|c|c|c|c|c|c|c|}
	\hline
	Dataset          & Entities & Relations &  Train Edges & Test Edges \\ \hline
    \multicolumn{5}{|c|}{Full KG}   \\ \hline
	{\tt MetaQA} & 43,234   & 18     & 150,160  & 8,000 \\ \hline
	{\tt CWQ} & 81,272   & 338       & 423,663 & 35,558  \\ \hline
	{\tt GQs} & 64,625   & 715       & 70,291  &  14,015 \\ \hline
    \multicolumn{5}{|c|}{50\% KG}   \\ \hline
    {\tt MetaQA} & 43,234   & 18     & 66,791  & 4,000 \\ \hline
	{\tt CWQ} & 73,821   & 292       & 245,876 &  35,558  \\ \hline
	{\tt GQs} & 64,625   & 715       & 35,145  &  14,059 \\ \hline
	\end{tabular}
 }
	\label{datasets}
\end{table}

\section{Ablation Study}
Table ~\ref{simple} shows the performance of simply combining LLMs with QTO. 
As we can see, the proposed method has better performance. 

\begin{table}[H]
	\centering
	\caption{Ablation Study of Combination of LLMs with KGR.}
	\setlength\tabcolsep{2pt}
 \scalebox{0.75}{
	\begin{tabular}{|c|c|c|c|c|c|c|}
	\hline
	Model          & \multicolumn{3}{c|}{chatGPT + QTO}  & \multicolumn{3}{c|}{\myModel}  \\ \hline
 \multicolumn{1}{|c|}{} & \multicolumn{6}{c|}{Full KG} \\ \hline
  Metric         & Hit@1 & Hit@3  & Hit@10 & Hit@1 & Hit@3  & Hit@10  \\ \hline
	MetaQA 1hop &  97.7 &  100.0  &  100.0  &  98.2  &  100.0  & 100.0 \\ \hline
	MetaQA 2hop  & 89.1  & 100.0   & 100.0   &  89.0  & 100.0 &  100.0    \\ \hline
	MetaQA 3hop &  90.3 & 98.9  &  99.7 &  89.0 & 99.3 & 99.7 \\ \hline
	GQ 2hop  &  100.0  & 100.0 & 100.0 &  100.0  & 100.0 & 100.0  \\ \hline
    GQ 2intersection  & 100.0  & 100.0 & 100.0 &  100.0  & 100.0 & 100.0   \\ \hline
    CWQ 3hop & 96.3  & 96.3  & 100.0  &  96.5  & 96.5  &  100.0  \\ \hline
    \multicolumn{1}{|c|}{} & \multicolumn{6}{c|}{50\% KG} \\ \hline
  Metric         & Hit@1 & Hit@3  & Hit@10 & Hit@1 & Hit@3  & Hit@10  \\ \hline
	MetaQA 1hop & 76.5  &  84.3 &   87.3 & \cellcolor{lightgray} 77.7  & \cellcolor{lightgray} 86.3   & \cellcolor{lightgray} 89.7  \\ \hline
	MetaQA 2hop  & 47.5  & 59.7  & 69.7  &  \cellcolor{lightgray} 58.7 & \cellcolor{lightgray} 79.0 & \cellcolor{lightgray} 83.7  \\ \hline    
	MetaQA 3hop & 60.2 & 73.4  & 83.5 & \cellcolor{lightgray} 67.9 & \cellcolor{lightgray} 83.9  & \cellcolor{lightgray} 90.8  \\ \hline
	GQ 2hop  & 49.6 & 50.4 &  \cellcolor{lightgray}55.0 &   \cellcolor{lightgray}53.8  & \cellcolor{lightgray}54.5  & 54.5   \\ \hline
    GQ 2intersection  &  44.4 &   44.4  &  44.4 &  44.4 &   44.4  &  44.4   \\ \hline
    CWQ 3hop &  27.6 & 37.9  &  48.2  &  \cellcolor{lightgray}33.3 &  \cellcolor{lightgray}40.7  &  \cellcolor{lightgray}59.2 \\ \hline
    
	\end{tabular}
 }
	\label{simple}
	\vspace{-1\baselineskip}
\end{table}

\section{The LLM Prompt for Projection Transformation}

In this paper, we use the following prompt for The Projection Transformation. 

\begin{myquote}
Prompt: Projection Transformation

\begin{inside_myquote}\small
I will give you a question. Please output at most 10 answers for this question in a json format. The output json has only key "answer entity".

Examples:

Q: movie interstellar starred by who?

A: {"answer entity": ["Matthew McConaughey", "Anne Hathaway", "Jessica Chastain", "Bill Irwin", "Ellen Burstyn", "Michael Caine"]}

Q: Which movie starred Tom Hanks?

A: {"answer entity": ["Forrest Gump", "A Man Called Otto", "Cast Away", "The Green Mile", "The Terminal, Saving Private Ryan"]}

Q: "\{text\_A\}"

A: 
\end{inside_myquote}

\end{myquote}

\section{Details for Backbone Fuzzy Logic Reasoning}\label{app:backbone-KGQA}

When constructing the relation matrix, we adopt the approach presented in~\cite{bai2023answering}. We define a neural adjacency matrix $M_r \in [0, 1]^{|\mathcal{E}|\times|\mathcal{E}|}$ for each relation $r$ as follows: $M_r[i,j] = r(e_i, e_j)$, where $e_i, e_j \in \mathcal{V}$ represent the $i$-th and $j$-th entities, respectively.
We use $g_{r_i}(e_i, e_j)$ to score the likelihood of $e_i$ is connected with $e_j$ by relation $r_i$, and we need to calibrate this score to a probability within the range [0, 1]. 
The calibration function must be monotonically increasing, and when there is an edge $r_i$ between $e_i$ and $e_j$ in the knowledge graph, the score should be accurately calibrated to 1.
Considering that there might be multiple valid tail entities for the triplet $(h, r, t)$, and they should all receive a probability close to 1, we multiply the normalized probability by the number of tail entities $N_t = \max\{1, |\{(e_i, r, e) \in \mathcal{L} | e \in \mathcal{V}\}|\}$. {It is} important to note that we set the lower bound to 1 to allow positive scores for long-tailed entity $e_i$ that does not have known edges of a certain type $r$ in the training set. Thus, we obtain:
\begin{align*}
    r(e_i, e_j) = \frac{\mathrm{exp}(g_r(h, t)) N_t}{\sum_{e \in \mathcal{V}} \mathrm{exp}(g_r(h, e))}
\end{align*}
We further round $r(e_i, e_j)$ so that it falls within the range [0, 1] and remains consistent with real triplets in the knowledge graph:
\begin{align*}
r(e_i, e_j) =
\begin{cases}
    1, \text{if } (e_i, r, e_j) \in \mathcal{L}  \\
    \min\{r(e_i, e_j), 1 - \delta\},  \text{otherwise} 
\end{cases}
\end{align*}

Here, we set $\delta = 0.0001 > 0$ to prevent overconfidence in predicted edges. It is worth noting that $M_r$ degenerates into an adjacency matrix when all entries less than 1 are set to 0.


\section{Embedding-based KGQA}\label{app:embedding}

The key to utilizing embedding-based KGQA methods, such as Q2B~\cite{ren2020query2box}, is very intuitive and simple. 

\noindent\textbf{(1) Computing the fuzzy vector.}
Given any query $q$, the embedding-based KGQA computes the embedding of this query $\vec{q}$. For each entity $e$, the score ${\rm Score}(e)$ can be computed by comparing $\vec{q}$ against the entity embedding $\vec{e}$. Then, the scores can be normalized into $[0,1]$ so that the fuzzy vector can be computed.

\noindent\textbf{(2) Combine LLM outputs into embeddings.}
As introduced in the main paper, the fuzzy vector will be updated by the outputs of LLMs in the stage of combination. Then, the key challenge is how to modify the query embedding $\vec{q}$ accordingly.

We consider the following linear correction
\begin{align}
    \vec{q}' = \vec{q} + \alpha \sum_{e_i\in \Delta} {\rm Score}(e_i) \vec{e_i},
\end{align}
where $\alpha$ is the hyperparameter, $\Delta$ is the entities whose scores are modified by LLMs, $\vec{e_i}$ is the entity embedding, and ${\rm Score}(e_i)$ is the LLM modified score after the combination.

Given that all embedding methods are demonstrated to be sub-optimal as QTO~\cite{bai2023answering}, we didn't test the embedding-based KGQA methods in our paper but left the way of correction for reference in future study.

\section{Examples of Atom Predicate to Question Prompt}
Table ~\ref{path_decoder} gives several examples of transforming knowledge graph relations to their corresponding question prompts. As we can see, the question prompts flows naturally are quite reasonable. 
\begin{table}[H]
	\centering
	\caption{Examples of relation to question prompt.}
	\small
 \scalebox{0.99}{
	\begin{tabular}{|c|c|c|}
	\hline
	\textbf{Relation}       & \textbf{Generated Query Template}    \\ \hline
	location.location.containedby & What is the location that contains entity1?  \\ \hline
    location.location.timeZones  & What is the time zone of entity1?  \\ \hline
    location.countyPlace.county & What is the county of entity1?  \\ \hline
    people.person.nationality  &  What is the nationality of entity1?  \\ \hline
    people.person.placeOfBirth & Where was entity1 born?   \\ \hline
    people.Relationship.sibling &  Who is the sibling of entity1?   \\ \hline
	\end{tabular}
 }
	\label{path_decoder}
	\vspace{-1\baselineskip}
\end{table}

\section{Examples of LARK Outputs}

\begin{table}[H]
    \centering
    \caption{Examples of LARK Outputs.}
    \small
    \scalebox{0.99}{
        \begin{tabular}{|p{4cm}|p{10cm}|} 
            \hline
            {Question} & 
            \texttt{Given the following (h,r,t) triplets where entity h is related to entity t by relation r;(5705,6,5706), (5706,7,5705), (5706,7,2641), (5706,7,7075). Answer the question: Which entities are connected to 5705 by relation 6? Return only the answer entities separated by commas with no other text.} \\ \hline
            {Answer} & \texttt{textbackslash begin code SELECT DISTINCT h FROM (SELECT h,r,t FROM (... (truncated outpu...))} \\ \hline
        \end{tabular}
    }
    \vspace{-1\baselineskip}
\end{table}

%% file: neurips_2024.bbl
\begin{thebibliography}{10}

\bibitem{gpt3}
Tom Brown, Benjamin Mann, Nick Ryder, Melanie Subbiah, Jared~D Kaplan, Prafulla Dhariwal, Arvind Neelakantan, Pranav Shyam, and Sastry.
\newblock Language models are few-shot learners.
\newblock In {\em Advances in Neural Information Processing Systems}, 2020.

\bibitem{bahdanau2016neural}
Dzmitry Bahdanau, Kyunghyun Cho, and Yoshua Bengio.
\newblock Neural machine translation by jointly learning to align and translate.
\newblock In {\em International Conference on Learning Representations}, 2016.

\bibitem{txt_generation}
Ilya Sutskever, James Martens, and Geoffrey Hinton.
\newblock Generating text with recurrent neural networks.
\newblock In {\em Proceedings of the 28th International Conference on International Conference on Machine Learning}, ICML'11. Omnipress, 2011.

\bibitem{recommendation}
Boxin Du, Lihui Liu, and Hanghang Tong.
\newblock Sylvester tensor equation for multi-way association.
\newblock In {\em Proceedings of the 27th ACM SIGKDD Conference on Knowledge Discovery and Data Mining}, KDD '21, New York, NY, USA, 2021. Association for Computing Machinery.

\bibitem{openai2023gpt4}
OpenAI.
\newblock Gpt-4 technical report, 2023.

\bibitem{llama2}
Hugo Touvron and Louis Martin.
\newblock Llama 2: Open foundation and fine-tuned chat models, 2023.

\bibitem{petroni2019language}
Fabio Petroni, Tim Rocktäschel, Patrick Lewis, Anton Bakhtin, Yuxiang Wu, Alexander~H. Miller, and Sebastian Riedel.
\newblock Language models as knowledge bases?, 2019.

\bibitem{pan2023unifying}
Shirui Pan, Linhao Luo, Yufei Wang, Chen Chen, Jiapu Wang, and Xindong Wu.
\newblock Unifying large language models and knowledge graphs: A roadmap, 2023.

\bibitem{wei2023chainofthought}
Jason Wei, Xuezhi Wang, Dale Schuurmans, Maarten Bosma, Fei Xia, Ed~Chi, Quoc~V Le, Denny Zhou, et~al.
\newblock Chain-of-thought prompting elicits reasoning in large language models.
\newblock {\em Advances in Neural Information Processing Systems}, 2022.

\bibitem{yao2023tree}
Shunyu Yao, Dian Yu, Jeffrey Zhao, Izhak Shafran, Thomas~L. Griffiths, Yuan Cao, and Karthik Narasimhan.
\newblock Tree of thoughts: Deliberate problem solving with large language models.
\newblock {\em Advances in Neural Information Processing Systems}, 2023.

\bibitem{besta2023graph}
Maciej Besta, Nils Blach, Ales Kubicek, Robert Gerstenberger, Lukas Gianinazzi, Joanna Gajda, Tomasz Lehmann, Michal Podstawski, Hubert Niewiadomski, Piotr Nyczyk, and Torsten Hoefler.
\newblock Graph of thoughts: Solving elaborate problems with large language models, 2023.

\bibitem{wang2021benchmarking}
Zihao Wang, Hang Yin, and Yangqiu Song.
\newblock Benchmarking the combinatorial generalizability of complex query answering on knowledge graphs.
\newblock In {\em NeurIPS Datasets and Benchmarks Track}, 2021.

\bibitem{yin2024rethinking}
Hang Yin, Zihao Wang, and Yangqiu Song.
\newblock Rethinking complex queries on knowledge graphs with neural link predictors.
\newblock In {\em The Twelfth International Conference on Learning Representations}, 2024.

\bibitem{ren2020query2box}
Hongyu Ren, Weihua Hu, and Jure Leskovec.
\newblock Query2box: Reasoning over knowledge graphs in vector space using box embeddings.
\newblock In {\em International Conference on Learning Representations}, 2020.

\bibitem{bai2023answering}
Yushi Bai, Xin Lv, Juanzi Li, and Lei Hou.
\newblock Answering complex logical queries on knowledge graphs via query computation tree optimization.
\newblock In {\em Proceedings of the 40th International Conference on Machine Learning}, 2023.

\bibitem{CQD}
Pasquale Minervini, Erik Arakelyan, Daniel Daza, and Michael Cochez.
\newblock Complex query answering with neural link predictors.
\newblock In {\em International Conference on Learning Representations}, pages 1--14, 2021.

\bibitem{chen2022fuzzy}
Xuelu Chen, Ziniu Hu, and Yizhou Sun.
\newblock Fuzzy logic based logical query answering on knowledge graphs.
\newblock In {\em Proceedings of the AAAI Conference on Artificial Intelligence}, 2022.

\bibitem{choudhary2024complex}
Nurendra Choudhary and Chandan~K. Reddy.
\newblock Complex logical reasoning over knowledge graphs using large language models, 2024.

\bibitem{newlook}
Lihui Liu, Boxin Du, Heng Ji, ChengXiang Zhai, and Hanghang Tong.
\newblock Neural-answering logical queries on knowledge graphs.
\newblock In {\em Proceedings of the 27th ACM SIGKDD Conference on Knowledge Discovery and Data Mining}, KDD '21, 2021.

\bibitem{ren2020beta}
Hongyu Ren and Jure Leskovec.
\newblock Beta embeddings for multi-hop logical reasoning in knowledge graphs.
\newblock {\em Advances in Neural Information Processing Systems}, 2020.

\bibitem{complEx}
Th\'{e}o Trouillon, Johannes Welbl, Sebastian Riedel, \'{E}ric Gaussier, and Guillaume Bouchard.
\newblock Complex embeddings for simple link prediction.
\newblock In {\em Proceedings of the 33rd International Conference on International Conference on Machine Learning - Volume 48}, ICML'16. JMLR.org, 2016.

\bibitem{king1998unifying}
Gary King.
\newblock {\em Unifying political methodology: The likelihood theory of statistical inference}.
\newblock University of Michigan Press, 1998.

\bibitem{robinson2023leveraging}
Joshua Robinson and David Wingate.
\newblock Leveraging large language models for multiple choice question answering.
\newblock In {\em The Eleventh International Conference on Learning Representations}, 2023.

\bibitem{embedkgqa}
Apoorv Saxena, Aditay Tripathi, and Partha Talukdar.
\newblock Improving multi-hop question answering over knowledge graphs using knowledge base embeddings.
\newblock In {\em Proceedings of the 58th Annual Meeting of the Association for Computational Linguistics}, pages 4498--4507, 2020.

\bibitem{lan:acl2020}
Yunshi Lan and Jing Jiang.
\newblock Query graph generation for answering multi-hop complex questions from knowledge bases.
\newblock In {\em Proceedings of the 58th Annual Meeting of the Association for Computational Linguistics (ACL)}, 2020.

\bibitem{graphQuestions}
Yu~Su, Huan Sun, Brian Sadler, Mudhakar Srivatsa, Zenghui Yan, and Xifeng Yan.
\newblock On generating characteristic-rich question sets for {QA} evaluation.
\newblock In {\em Proceedings of the 2016 Conference on Empirical Methods in Natural Language Processing}. Association for Computational Linguistics, 2016.

\bibitem{gpt2}
Alec Radford, Jeffrey Wu, Rewon Child, David Luan, Dario Amodei, Ilya Sutskever, et~al.
\newblock Language models are unsupervised multitask learners.
\newblock {\em OpenAI blog}, 2019.

\bibitem{touvron2023llama}
Hugo Touvron and Thibaut Lavril.
\newblock Llama: Open and efficient foundation language models, 2023.

\bibitem{manyika2023overview}
James Manyika.
\newblock An overview of bard: an early experiment with generative ai, 2023.

\bibitem{kojima2023large}
Takeshi Kojima, Shixiang~(Shane) Gu, Machel Reid, Yutaka Matsuo, and Yusuke Iwasawa.
\newblock Large language models are zero-shot reasoners.
\newblock In {\em Advances in Neural Information Processing Systems}, 2022.

\bibitem{zhou2023leasttomost}
Denny Zhou, Nathanael Sch{\"a}rli, Le~Hou, Jason Wei, Nathan Scales, Xuezhi Wang, Dale Schuurmans, Claire Cui, Olivier Bousquet, Quoc~V Le, and Ed~H. Chi.
\newblock Least-to-most prompting enables complex reasoning in large language models.
\newblock In {\em The Eleventh International Conference on Learning Representations}, 2023.

\bibitem{creswell2022selectioninference}
Antonia Creswell, Murray Shanahan, and Irina Higgins.
\newblock Selection-inference: Exploiting large language models for interpretable logical reasoning.
\newblock In {\em The Eleventh International Conference on Learning Representations}, 2023.

\bibitem{luo2024reasoning}
Linhao Luo, Yuan-Fang Li, Gholamreza Haffari, and Shirui Pan.
\newblock Reasoning on graphs: Faithful and interpretable large language model reasoning, 2024.

\bibitem{sun2024thinkongraph}
Jiashuo Sun, Chengjin Xu, Lumingyuan Tang, Saizhuo Wang, Chen Lin, Yeyun Gong, Lionel~M. Ni, Heung-Yeung Shum, and Jian Guo.
\newblock Think-on-graph: Deep and responsible reasoning of large language model on knowledge graph, 2024.

\bibitem{lewis2021retrievalaugmented}
Patrick Lewis, Ethan Perez, and Aleksandra Piktus.
\newblock Retrieval-augmented generation for knowledge-intensive nlp tasks, 2021.

\bibitem{edge2024local}
Darren Edge, Ha~Trinh, Newman Cheng, Joshua Bradley, Alex Chao, Apurva Mody, Steven Truitt, and Jonathan Larson.
\newblock From local to global: A graph rag approach to query-focused summarization, 2024.

\bibitem{shi2023replug}
Weijia Shi, Sewon Min, and Michihiro Yasunaga.
\newblock Replug: Retrieval-augmented black-box language models, 2023.

\bibitem{zhu2022neuralsymbolic}
Zhaocheng Zhu, Mikhail Galkin, Zuobai Zhang, and Jian Tang.
\newblock Neural-symbolic models for logical queries on knowledge graphs.
\newblock In {\em International Conference on Machine Learning}. PMLR, 2022.

\bibitem{Choudhary_2021}
Nurendra Choudhary, Nikhil Rao, Sumeet Katariya, Karthik Subbian, and Chandan~K. Reddy.
\newblock Self-supervised hyperboloid representations from logical queries over knowledge graphs.
\newblock In {\em Proceedings of the Web Conference 2021}. {ACM}, 2021.

\end{thebibliography}
